\documentstyle[pre,aps,amsfonts,floats,epsfig]{revtex}

\begin{document}


\twocolumn[\hsize\textwidth\columnwidth\hsize\csname@twocolumnfalse%
\endcsname

\title {Cooling dynamics of a dilute gas of inelastic rods: a
        many particle simulation}
\author{Timo Aspelmeier, G\"otz Giese and 
        Annette Zippelius}

\address{
Institut f\"ur Theoretische Physik,
Universit\"at G\"ottingen,      \\
Bunsenstrasse 9, D-37073 G\"ottingen, Germany}

\date{\today}

\maketitle

\begin{abstract}
  We present results of simulations for a dilute gas of inelastically
  colliding particles. Collisions are modelled as a stochastic process, which
  on average decreases the translational energy (cooling), but allows for
  fluctuations in the transfer of energy to internal vibrations. We show that
  these fluctuations are strong enough to suppress inelastic collapse. This
  allows us to study large systems for long times in the truely inelastic
  regime. During the cooling stage we observe complex cluster dynamics, as
  large clusters of particles form, collide and merge or dissolve. Typical
  clusters are found to survive long enough to establish local equilibrium
  within a cluster, but not among different clusters. We extend the model to
  include net dissipation of energy by damping of the internal vibrations.
  Inelatic collapse is avoided also in this case but in contrast to the
  conservative system the translational energy decays according to the mean
  field scaling law, $E(t)\propto t^{-2}$, for asymptotically long times.

\vspace{1cm}
\end{abstract}
]
\section{Introduction}

In a recent paper \cite{GieseAZ96}, hereafter referred to as I, we discussed
the properties of inelastic \emph{two particle} collisions, starting from a
Hamiltonian model for one dimensional elastic rods. Within this model, the
coefficient of restitution $\epsilon$ does not only depend on the relative
velocity of the colliding particles, but in addition becomes a
\emph{stochastic} quantity, depending on the state of excitation of the
internal vibrations. Here we extend the analysis to a discussion of the
\emph{many body} dynamics of a one-dimensional gas of granular particles,
modelled as elastic rods. We concentrate on dilute granular systems in the
"grain inertia" regime, where two particle collisions dominate the dynamics.
It was shown in I that successive collisions are to a very good approximation
uncorrelated, so that the many body dynamics is a random Markov process.
Consequently, the collisions are simulated by a Monte Carlo algorithm:
velocities are updated with a random coefficient of restitution, drawn from
the appropriate probability distribution. In between collision events,
particles move freely like in an event driven algorithm.
 
We focus here on the cooling properties of a large system (10000 particles) in
the inelastic regime and refer to cooling as the decay of
\emph{translational} energy with time. We observe the evolution of spatial
structures, without running into problems with inelastic collapse, which is
always avoided by the algorithm.  The most prominent spatial structures are
large clusters of particles, which are seen to form and \emph{decay} by
colliding with other clusters.  The velocity distribution within a cluster is
to a good approximation Maxwellian, whereas the global velocity distribution
shows significant deviations from Maxwellian, indicating that local
equilibrium has been established within a cluster, but not among different
clusters.

The model is extended to include net dissipation, i.e. irreversible energy
loss, in a phenomenological way by simply introducing a single relaxation time
for the decay of the energy of internal vibrations. The final state of this
model is one big cluster with all particles at rest. The dynamics with
dissipation resembles a deterministic system (i.~e. with constant coefficient
of restitution) as long as no inelastic collapse is threatening. When the
collision frequency increases dramatically, then ultimately the time between
collisions will become smaller than the decay time for the internal energy, so
that the vibrations no longer decay in between collisions. Then the model
effectively reduces to the above stochastic one without dissipation and no
inelastic collapse occurs. The kinetic energy of translation follows on
average the mean field scaling law $E(t) \propto t^{-2}$.

Several groups have simulated onedimensional granular media using event driven
algorithms with constant coefficient of restitution.  For $(1-\epsilon)\ge
const./N$ a divergence of the collision frequency in finite time, i.e.
inelastic collapse is observed.  This leads to a breakdown of the algorithm
and one either has to restrict oneself to the quasielastic regime, where
$\epsilon $ is sufficiently close to one or additional assumptions about the
dynamics of clusters have to be made.  Bernu and Mazighi \cite{Bernu90}
investigate a column of beads colliding with a wall. McNamara and Young
\cite{Young93} and Sela and Goldhirsch \cite{Sela95} discuss the cooling
dynamics of a granular gas in the quasielastic regime. They observe the
evolution of spatial structures and a bimodal velocity distribution. The
critical wavelength of the instability is related to the minimum number of
particles for inelstic collapse to occur, given a fixed value of $\epsilon$.
Clement et al. \cite{Clement93} and Luding et al. \cite{Luding94} study a
vertical column of beads in a gravitational field with a vibrating bottom
plate. For $\epsilon$ close to one they observe a fluidization transition,
whereas for $\epsilon \ll 1$ a bifurcation scenario is seen to
take place. The latter has also been observed by Luck et al. \cite{Luck93} for
a single bead on a vibrating plane.

All of the above simulations use a coefficient of restitution which is
independent of the impact velocity whereas experiments on ice spheres
reveal a velocity dependence of $\epsilon$ \cite{Bridges88}.
There have also been several attempts to calculate the velocity dependence
of the coefficient of restitution by extending the static theory of
Hertz \cite{Hertz1882} to viscoelastic behaviour. One either assumes a
phenomenological damping term \cite{Poeschl28} in the equation of
motion for the deformation or alternatively uses a quasistatic
approximation \cite{Pao55,Brilliantov96} for low relative impact
velocities. As a result of either
approximation the coefficient of restitution becomes velocity
dependent. Simulations of large systems with strongly inelastic
collisions have been performed with this model \cite{Spahn97}.

Another approach is based on phenomenological wave theory. Here one
assumes that two colliding bodies do not vibrate before collision and
that the impact triggers a travelling elastic wave in both of them.
For one dimensional rods this ansatz yields \cite{Auerbach}
$\epsilon = l_1/l_2 $, independent of the relative velocity of the
colliding particles. Here $l_1$ ($l_2$) denotes the length of the shorter
(longer) rod. As shown in I, these results are contained in
our model.

Our paper is organized as follows. In the next section we review the model of
I, discuss the probability distribution of $\epsilon$ and define the algorithm
for the many body dynamics. Results of simulations are presented in
sec. \ref{simulations}. We first discuss global quantities, like the time decay
of the kinetic energy and the total number of collisions as a function of
time. Subsequently we analyse the local structure with help of the
paircorrelation function and discuss the formation and decay of particle
clusters, as well as the distribution of the particles' velocities.  In
sec. \ref{damping} the model with dissipation of vibrational energy is
introduced. Finally in sec. \ref{conclusion} we summarize our results and give
an outlook to forthcoming work.

\section{Markovian dynamics of inelastic rods}

We first review the Hamiltonian model of I and summarize the 
properties of two particle collisions.
We then show that the transition probabilities of 
the resulting Markov process obey detailed balance and introduce
the algorithm for the dynamics of the many-body system. 

\subsection{Two-particle collisions}

Our starting point are the Hamiltonian equations of motion of a system
of $N$ elastic rods of homogeneous mass density. 
The particles are placed on a ring of circumference $L$.
Each rod is
characterized by its length $l_i$, total mass $m_i$ and centre of mass position
$R_i(t)$. Its vibrational excitations are described by
$N_{mod}$ normal modes $q_i^{\nu}  (\nu=1,\ldots,N_{mod})$ of wavenumber
$k_{i,\nu}=\pi \nu /l_i$ and frequency $\omega_{i,\nu}=ck_{i,\nu}$. 
The only important material parameter for our model is the sound velocity $c$.
 We model collisions of the rods by a
short range repulsive potential $V(r)=B\exp{(-\alpha r)}$, which depends
on the momentary end-to-end distance $r$ between the colliding rods, thus
coupling translational and vibrational degrees of freedom. We shall be
interested in the hard core limit, which can be achieved by letting
$\alpha \to \infty$ \footnote{The constant B is arbitrary, it can be
  absorbed by rescaling time $t\to t \sqrt{B}$ and frequencies $\omega
  \to \omega/\sqrt{B}$.}. The total Hamiltonian of our model reads:

\begin{eqnarray}
\label{Ges_Ham}
   {\cal H}&=&
{\cal H}_{bath}\left\{p_{i}^{(\nu)},q_{i}^{(\nu)}\right\} + 
{\cal H}_{tr}\left\{P_i\right\} +
{\cal H}_{int}\left\{R_i,q_i^{(\nu)}\right\}\nonumber\\
&=&\sum\limits_{i=1}^{N}\sum\limits_{\nu=1}^{N_{mod}}\left\{
\frac{{p_{i}^{(\nu)}}^{2}}{2 m_{i}}+m_{i}\omega_{i,\nu}^{2}
\frac{{q_{i}^{(\nu)}}^{2}}{2}\right\}
+\sum\limits_{i=1}^{N}\frac{P_{i}^{2}}{2 m_{i}} + \\
&&\sum\limits_{i=1}^{N-1}B
e^{-\alpha\left(R_{i+1,i}
+\sqrt{2}\sum\limits_{\nu}\left(
q_{i+1}^{(\nu)}-(-1)^{\nu}q_{i}^{(\nu)}\right)\right) }
\quad .\nonumber
\end{eqnarray}
Here $R_{i+1,i} = R_{i+1}-R_i-(l_{i+1}-l_i)/2$ is the end-to-end distance
of two undeformed neighbouring rods and $P_i(t)$
 and $p_i(t)$ denote the conjugate momenta for the centre
of mass and the amplitude of vibration.
 The first term, ${\cal H}_{bath}$,
models the internal vibrations, the second term, ${\cal H}_{tr}$, the
translational motion, and the third term, ${\cal H}_{int}$, the interaction
between the rods.

In I we have analysed the statistical properties of two-particle collisions by
numerically integrating the full Hamiltonian dynamics for the case $N=2$ with
length ratio $\gamma=l_1/l_2$. The main results are the following.
Equipartition among the vibrational states of a rod is achieved fast (after
about five collisions) as compared to the relaxation of the translational
velocity, which happens on a time scale of about 80 collisions. The
coefficient of restitution of two successive collisions is to a very good
approximation uncorrelated. Based on these observations a simplified
description was achieved in I by integrating out the internal degrees of
freedom, which can be done exactly for a single two particle collision. One is
left with an effective equation of motion for the rescaled relative velocity
of the two rods $w(\tau)= \dot{R}_{2,1}(t)/\dot{R}_{2,1}(0)-1$.
Here $\tau=ct/l$ and $l=2l_1l_2/(l_1+l_2)$ denotes an effective length
which is always chosen much larger than the range of
the potential, i.e. $\alpha l \gg 1$. 
Based on the observation of fast equipartition among the vibrational
states, these are modelled by a {\it thermalized bath}, characterized
by a temperature $T_B=E_{bath}/N_{mod}$, where the vibrational energy
of a rod is given by the sum of the energies of the individual modes:
$E_{bath}=\sum_{\nu=1}^{N_{mod}}E_{\nu}$. 
Thus $q_i^{\nu}(0)$ and $p_i^{\nu}(0)$ are
taken as independent, canonically distributed random variables
\begin{eqnarray}
\label{Q_Coeff}
  \langle q_i^{(\nu)}(0)\rangle&=&\langle p_i^{(\nu)}(0)\rangle\,\,=\,0
    \nonumber\\
\left\langle (q_i^{(\nu)}(0))^2\right\rangle&=&
  \left\langle
    \left(\frac{p_i^{(\nu)}(0)}{m_i\omega_{i,\nu}}\right)^2
  \right\rangle 
= \frac{T_B}{m_i\omega_{i,\nu}^2}.
\end{eqnarray}
Under these assumptions the relative velocity $w(\tau)$ obeys a
{\it stochastic} equation of motion
\begin{eqnarray}
  \label{SDE}
\frac{d}{d\tau}
w(\tau)&=& \frac{1}{\kappa}
\exp\bigg\{\kappa\big(\tau-\tau_0-w(\tau) - \nonumber\\
&& \quad \sum\limits_{i=1}^2
  \sum\limits_{n=1}^{\infty}w(\tau-n\Gamma_{i})
  +q(\tau)\big)\bigg\},
\end{eqnarray}
where $\kappa = -\frac{\alpha l}{c}\dot{R}_{2,1}(0)$ and
 $\tau_0=\alpha R_{2,1}(0)/\kappa$.
The coefficient of restitution is given by
\begin{equation}
\label{e:epsilon}
\epsilon = \lim_{\tau\to\infty}w(\tau)-1
\end{equation}
and thus becomes a {\it stochastic} variable,
depending on the state of the vibrational bath before the collision.
$q(\tau)$ is a Gaussian random noise with zero mean and covariance 
\begin{eqnarray}
\label{Cov_QQ}
C_q(\tau)&=&\langle q(\tau')q(\tau'+\tau)\rangle \nonumber\\
&=&\Biggl\{\sum_{i=1}^2\frac{1}{2\Gamma_i}
\Biggl(\tau-\frac{\Gamma_i}{2} - \Gamma_i\sum_{n=1}^{\infty}
\theta(\tau-n\Gamma_i)\Biggr)^2- \\
&& \quad\frac{\Gamma_1\Gamma_2}{24}\Biggr\}\sigma^2, \nonumber \\
\sigma ^2 & = & \frac{T_B}{2E_{tr}}. \nonumber
\end{eqnarray}
Here, $\theta$ denotes the Heaviside step function and
$E_{tr}=\mu(\dot{R}_{2,1}(0))^2/2$ is the translational energy of the colliding
rods in their center of mass frame of reference.  The $\Gamma_i$ which appear
in eqs. (\ref{SDE}) and (\ref{Cov_QQ}) are determined by the length ratio
$\gamma$ of the rods according to $\Gamma_1=1+\gamma$ and
$\Gamma_2=1+\frac{1}{\gamma}$ and $\mu=m_1m_2/(m_1+m_2)$ is the effective
mass. The stochastic process $\{q(\tau)\}$ is simply related to two periodic
brownian bridge processes \cite{Kloeden} with periods $\Gamma_1$ and
$\Gamma_2$ respectively.

In the hard core limit ($\kappa\to\infty$) the stochastic equation for the
rescaled relative velocity $w(\tau)$ (eq. (\ref{SDE})) can be solved by saddle
point methods, yielding
\begin{eqnarray}
\label{e:mysol}
w(\tau) &=& \max\left(0, f(\tau)\right),\quad\mbox{where}\nonumber \\
f(\tau) &=& \max_{0\leq\tau'\leq\tau}
  \biggl\{ \tau' - \tau_0 - \sum_{i,\nu}w(\tau'-\nu\Gamma_i) +
          q(\tau') \biggr\}.
\end{eqnarray}
The duration of the collision as well as the final velocity are
stochastic variables. The collision is ended when the memory terms in
eq. (\ref{e:mysol}) overcompensate the gain from the other terms 
$\tau' - \tau_0  +q(\tau') $, which are on average increasing. 

\subsection{Transition Probability}
\label{transition}

The results of the preceding section are interpreted as a Markov process in
discrete time, which accounts for transitions of the translational energy upon
successive collisions. During a collision $E_{tr}$ changes to a new value
$E'_{tr}=E_{tr} \epsilon^2$.  The probability for this transition is
determined by the probability density for the coefficient of restitution
$p_{\beta}(\epsilon)$ according to
\begin{eqnarray}
  \label{trans_dens}
   p_{\scriptscriptstyle T_B}(E_{tr}\rightarrow E'_{tr}) &=&  
  \frac{1}{2\epsilon E_{tr}}
p_{\beta}\!(\epsilon)
\Biggr|_{\epsilon=\sqrt{\frac{E'_{tr}}{E_{tr}}}}\quad, \nonumber\\
\beta &=& \frac{E_{tr}}{T_B}
\end{eqnarray}
($T_B$ here denotes the bath temperature of \emph{both} rods, under the
assumption that the temperatures are equal. If that is not the case, one would
have to replace the index $\beta$ by two indices $\beta_1$ and $\beta_2$.
Here, we use only one index for notational simplicity.)

Changes in the bath temperature are not independent, but determined by energy
conservation:
\begin{equation}
  \label{E_change}
T'_{B}=T_{B}+\frac{1-\epsilon^2}{2N_{mod}}E_{tr}
\end{equation}

The stationary state of the Markov process is known: after cooling, the
system of two particles, each equipped with an internal bath, evolves
into a stationary state with a Boltzmann distribution for $E_{tr}$
\begin{equation}
  \label{Etr_stat}
  p_{\scriptscriptstyle T_B^0}^{stat}(E_{tr})
=\frac{1}{T_B^0}\exp\left(-\frac{E_{tr}} 
{T_B^0}\right) 
\end{equation}
with the bath temperature $ T_B^0=\frac{E_{tot}}{2N_{mod}+1}$ where
$E_{tot}$ is the total energy of the system.

It can be proven \cite{tobepublished} that this collision process obeys
detailed balance, which gives the following relation for $p_{\beta}(\epsilon)$
(this relation only holds if the temperature of both rods is indeed equal,
i.~e. $\beta_1=\beta_2=\beta$):
\begin{equation}
p_{\beta}(\epsilon)e^{-\beta} = 
  p_{\epsilon^2\beta}\left(\frac{1}{\epsilon}\right)e^{-\epsilon^2\beta}
\end{equation}

When the temperatures of the baths of oscillators are zero\footnote{Actually,
  it is sufficient that only the temperature of the longer rod is zero.},
i.~e. $q(\tau) = 0$ for all $\tau$, the collision is deterministic. For this
case, the coefficient of restitution is equal to $\gamma$, the ratio of the
lengths of the rods (see I and \cite{Auerbach}). Thus in the limit of small
temperatures, $p_{\beta}(\epsilon)$ should approach a $\delta$-Function
centered around $\gamma$.

At large temperatures, on the other hand, simulations suggest the (quite
sensible) result that $p_{\beta}(\epsilon)$ is a uniform distribution, i.~e.\ 
all possible $\epsilon$'s are equally probable.

It can be shown from equation (\ref{e:mysol}) that for $\gamma=1$ the
collision is always deterministic, i.~e. $\epsilon=1$ for any realisation
of the stochastic process ${q(\tau)}$. This interesting result will have
implications on our setup of the simulations (see sec. \ref{algorithm}).

\subsection{Algorithm}
\label{algorithm}

We now consider the dynamic evolution of $N$ particles on a ring of
circumference $L+\sum_{i=1}^Nl_i$. $L$ is thus the total length of the
interparticle spacings. For the following arguments the actual lengths of the
particles are unimportant because the point in time when a collision occurs
depends only on the end-to-end distance and the outcome of a collision depends
only on the length ratio. In order to keep the notation simple we map the
system to an equivalent one consisting of $N$ point particles on a ring of
circumference $L$. Each particle is characterized by its position $R_i(t)$,
its velocity $\dot{R}_i(t)$, the temperature of its internal bath
$T_{B}^{(i)}(t)$. The $N_{mod}$ internal modes of one rod are
represented by one degree of freedom only, namely
$T_B^{(i)}=\sum_{\nu=1}^{N_{mod}} E_i^{\nu}/N_{mod}$.
The rods are assigned alternating lengths such that the ratio $\gamma$ for each
collision has a fixed value, in our case $0.8$.
The ratio of masses
is also given by $\gamma$, assuming the same homogeneous mass density for both
kinds of rods. We choose rods of alternating length because due to the result
given at the end of sec. \ref{transition} a length ratio of $\gamma=1$ implies
$\epsilon=1$ always which would correspond simply to a standard one
dimensional hard sphere gas.

The model we use is a hybrid of an event
driven algorithm and a Monte Carlo simulation. The particles move freely in
between collisions, as in event driven algorithms.  When two particles collide
their states are updated stochastically, according to the distribution of
$\epsilon$.

It is convenient to introduce dimensionless variables $x_i=R_i N/L$ and
$v_i=\dot{R}_i \sqrt{\mu/T_0}$. $T_0$ serves as an energy scale and will be
identified with the homogeneous initial granular temperature of the many
particle system. Time is measured in units of $L\sqrt{\mu/T_0}/N$.

For the algorithm we only need relative distances and velocities
\begin{eqnarray}
\Delta x_i &=& \biggl\{
\begin{array}{rcl}
x_{i+1}-x_i & \mbox{for} & 1 \le i \le N-1 \\
N+x_1-x_N   & \mbox{for} & i=N
\end{array} \\
\Delta v_i &=& \biggl\{
\begin{array}{rcl}
v_{i+1}-v_i & \mbox{for} & 1 \le i \le N-1 \\
v_1-v_N   & \mbox{for} & i=N
\end{array}
\end{eqnarray}
The algorithm is defined by iteration of the following steps:

\begin{enumerate}
\item Calculate the time difference $\Delta t$ for the next collision to take
  place:
\begin{equation}
  \Delta t=\min_{\{i|\Delta v_i<0\}}\left(-\frac{\Delta x_i}{\Delta
      v_i}\right)
\end{equation}
The pair of particles which is going to collide next is denoted by
$(i_0,i_0+1)$.

\item The relative distances of all particles are updated according to:
\begin{equation}
  \Delta x_i(t+\Delta t)=\Delta x_i(t)+\Delta v_i(t)\Delta t.
\end{equation}
For the designated pair $(i_0,i_0+1)$ we obtain $\Delta x_{i_0}(t+\Delta
t)=0$.

\item The kinetic energy of relative motion of the pair $(i_0,i_0+1)$ as well
  as the mean local bath temperature are calculated according to
  $E_{tr}=\Delta v^2_{i_0}/2$ and $T_B=(T_B^{(i_0)}+T_B^{(i_0+1)})/2$.
  Subsequently, a random value of $\epsilon$ is chosen from the probability
  distribution $p_{\beta}(\epsilon)$, presently calculated by numerically
  solving eq. (\ref{e:mysol}) and applying eq. (\ref{e:epsilon}).
  
\item The bath temperatures and relative velocities are updated
\begin{eqnarray}
  T_B^{(i_0)}(t+\Delta t) & = & T_B^{(i_0+1)}(t+\Delta t) \nonumber\\
  &=&T_B+\frac{1-\epsilon^2}
  {2N_{mod}}E_{tr}\\
  \Delta v_{i_0-1}(t+\Delta t) & = & \Delta v_{i_0-1}(t)+\frac{1+\epsilon}{2}
  \Delta v_{i_0}(t) \\
  \Delta v_{i_0}(t+\Delta t) & = & -\epsilon \Delta v_{i_0}(t)\\
  \Delta v_{i_0+1}(t+\Delta t) & = & \Delta v_{i_0+1}(t)+\frac{1+\epsilon}{2}
  \Delta v_{i_0}(t).
\end{eqnarray}

\item Continue with step 1.

\end{enumerate}

\section{Simulations}
\label{simulations}

Many-body simulations using the above algorithm have been performed to
study the cooling dynamics of the system. More precisely, we focus
here on the intermediate range of timescales where equipartition among
the internal modes has already been achieved and the final equilibrium
state of equipartition among all degrees of freedom is not yet
reached. As shown below this time range extends over several orders of
magnitude.

We assume that two particle collisions dominate the dynamic evolution of the
system. This is justified for a dilute granular gas. The typical time of
interaction in our model is given by $t_{int}=2l/c$, i.e. the time a signal
needs to travel back and forth on a rod. Hence in principal, two colliding
rods can interact with a third one. This will be highly unlikely, as long as
the time between collisions is much longer than $t_{int}$. This requires $2l/c
\ll L/(N|\dot{R}_{i+1,i}|)$. So either the length of the rods has to be chosen
sufficiently small as compared to the mean distance $L/N$ or the initial
velocities $|\dot{R}_{i+1,i}|$ should be small compared to the velocity of
sound. The latter is a material parameter and can have quite high values for
hard materials (e.g. for steel, $c\sim 10^4 m/s$), favouring short interaction
times. In a standard event-driven simulation inelastic collapse occurs when
the number of particles is sufficiently large, resulting in a diverging
collision frequency.  This would clearly violate the condition that the time
between two collisions is long compared to $t_{int}$. However, since our
algorithm avoids the inelastic collapse, as will be discussed below, we will
still make use of the assumption that three or more particle collisions will
not be important.

A system of 10000 particles has been simulated and for most part a length
ratio $\gamma = 0.8$ has been used.  We start from a spatially homogeneous
distribution of particles having a maxwellian velocity distribution with
$\langle \frac{\Gamma_j}{2}v_j^2\rangle=\frac{1}{2}$ ($j=1 (2)$ stands for
the shorter (longer) species of rods). We use $N_{mod}=1000$
vibrational modes per particle. Initially the internal bath temperature
$T_B^{(i)}$ of each particle is set to 0.  

Our simulations were performed on a cluster of Linux workstations with
Pentium processors. The longest runs took about three weeks of computer
time.


\subsection{Global quantities}

\noindent\textbf{Kinetic energy}

\begin{figure}[htb]
\epsfig{file=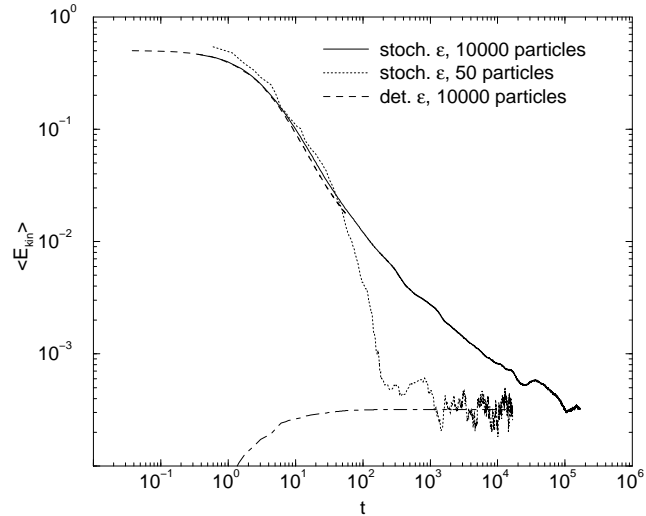,width=8.5cm}
\caption{Reduction of the kinetic energy per particle as a function of time.
The curve for the deterministic coefficient of restitution breaks off because
an inelastic collapse occured. The dot-dashed line shows the average
energy per vibrational mode for the 50 particle run and illustrates
that equipartition holds in the stationary state.}
\label{fig:cool1}
\end{figure}

\begin{figure}[htb]
\epsfig{file=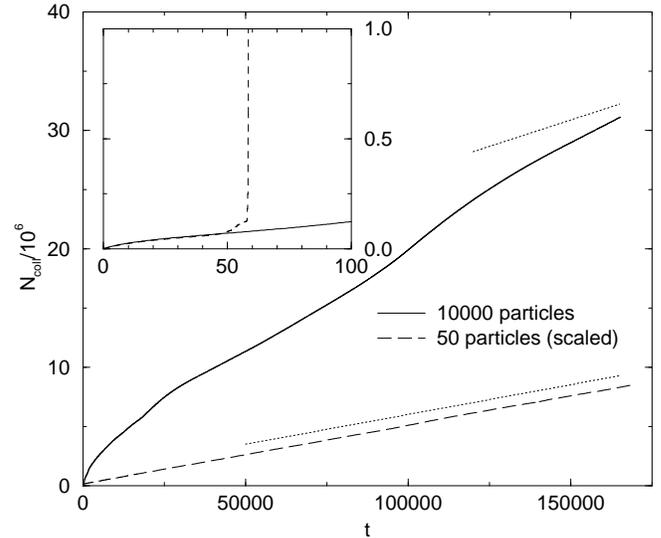,width=8.5cm}
\caption{Number of collisions as a function of time. The inset shows a 
comparison of the deterministic model (dashed line) and the stochastic
model (solid line). (The units on the axes of the inset are the same as on the
regular axes.)
The deterministic model quickly runs into the inelastic collapse,
seen by the diverging number of collisions. The dotted lines shows the
theoretical number of collisions as a function of time in the stationary
state according to eq. (\ref{coll.rate}) for the 10000 and the 50 particle
runs. The data for the 50 particle run has been scaled by a factor of 100
in order to fit on the graph.}
\label{fig:coll1}
\end{figure}

The time development of the total kinetic energy, which is given by
\begin{equation}
E_{kin}=\sum_{i=1}^N \frac{\Gamma_i}{2}v_i^2
\end{equation}
in our rescaled units,
is shown in fig. \ref{fig:cool1}, in comparison with results for the
deterministic model with constant $\epsilon$.
For small times the curves for the deterministic and the stochastic dynamics
are rather similar. In the initial stage very little energy is stored in the
internal modes and hence the coefficient of restitution is approximately given
by the deterministic value.  However the deterministic dynamics runs very
quickly into the inelastic collapse, as can be seen from the total number of
collisions which is shown as a function of time in fig. \ref{fig:coll1}.  When
this happens, the simulation gets stuck
so that the curve for the kinetic energy breaks off in fig. \ref{fig:cool1}. 
The stochastic dynamics shows completely different behaviour: The kinetic
energy continues to decrease until equilibrium is reached, where $E_{kin}$
continues to fluctuate around the stationary value which is given by
$E_{kin}^{stat}=E_{kin}(t=0)/(2N_{mod}+1)$. (The final state has not quite
been reached for the 10000 particle run in the time interval that is shown in
fig. \ref{fig:cool1}.)

The final state of our stochastic
model is a consequence of the idealised assumption that the total system be
conservative. In a more realistic model of granular media one expects the
particles to be at rest in the final state. In sec. \ref{damping} we shall
present a phenomenological extension of our model, which does take into
account energy dissipation of the microscopic degrees of freedom. In the final
state of the model with energy dissipation {\it all} particles are at rest.

\vspace{1ex}
\noindent\textbf{Collision rate}

Simple mean field arguments \cite{Sela95} have been used to derive scaling
laws for the time evolution of kinetic energy and collision rate. One assumes
that the particle velocities are uncorrelated and Gaussian distributed. For a
constant coefficient of restitution one obtains $E_{kin}(t) \sim t^{-2}$ and
$\dot{N}_{coll} \sim \ln t$. Neither scaling law fits our data, as can be seen
from figs. \ref{fig:cool1} and \ref{fig:coll1}. McNamara and Young \cite
{Young93} have already pointed out that the mean field scaling laws are only
applicable in the quasi elastic regime, where no inelastic collapse occurs.
Otherwise the assumption of uncorrelated Gaussian velocities breaks down. In
the stochastic model we have additional fluctuations of the coefficient of
restitution, which invalidate the derivation of the above scaling laws. Hence
it is no surprise that the data disagree with these relations.

The rate of collisions becomes constant as the stationary state is approached,
as can be seen from fig. \ref{fig:coll1}. The average collision rate is given
by $\dot{N}_{coll}=N\overline{\Delta v}/(2\overline{\Delta x})$.  In the
stationary state the velocities are indeed uncorrelated Gaussian variables,
distributed according to
\begin{equation}
  p_j(v)=\left(\frac{2\pi}{2N_{mod}+1}\right)^{-1/2}
  \exp{\left(-\frac{\Gamma_j v^2}{2(2N_{mod}+1)^{-1}}\right)},
\end{equation}
where $j=1(2)$ again stands for the shorter (longer) type of rods.
We assume $\overline{\Delta x} =1$ and perform the average over velocities to
obtain
\begin{equation}
\label{coll.rate}
\dot{N}_{coll}^{stat}=\frac{N}{2\sqrt{\pi(2N_{mod}+1)}}.
\end{equation}
This result is in very good agreement with the simulations of the 50 particle
system in the stationary state (see fig. \ref{fig:coll1}). The 10000 particle
system is also approaching the correct value as it gets closer to the
stationary state.

\subsection{Local quantities}


\noindent\textbf{Particle density}
\label{particle.density}
\begin{figure}[hbt]
\epsfig{file=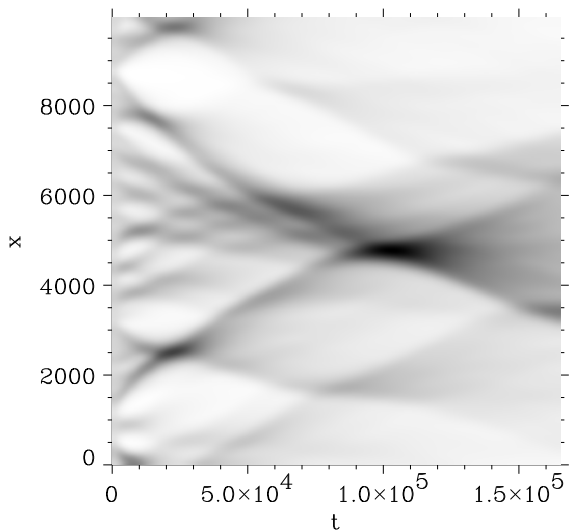,width=8.5cm}
\caption{Time evolution of the particle density. Dark regions
  indicate high density.}
\label{fig:dichte1}
\end{figure}
\begin{figure}[htb]
\epsfig{file=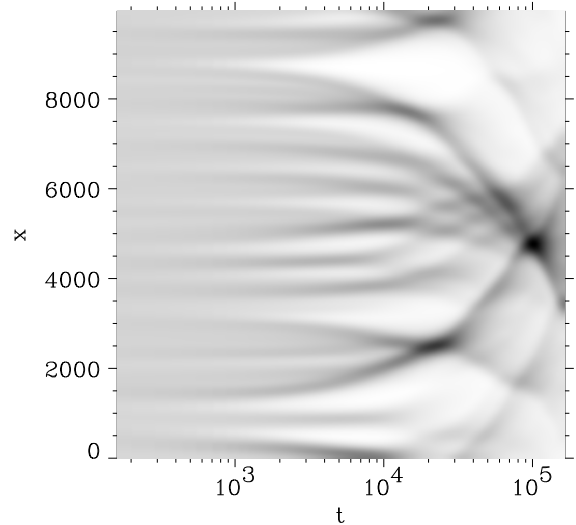,width=8.5cm}
\caption{The same as fig. \ref{fig:dichte1} on a logarithmic time scale}
\label{fig:dichte2}
\end{figure}

As is well known, inelastic particles without internal structure tend to
cluster in one dimension. This clustering leads to a break down (inelastic
collapse) of the system if $\epsilon$ is constant and less than a critical
value that depends solely on the number of particles. In our model particles
have internal degrees of freedom and the translational energy is not
completely lost in a collision but is stored in the internal vibrations and
can be transferred back to the translational motion. Due to the properties of
$p_{\beta}(\epsilon)$ (see section \ref{transition}), the probability for this
to happen gets larger as the translational energy decreases. Therefore
clusters do form but dissolve after a while and no inelastic collapse takes
place.

We start again from a spatially homogeneous distribution of
particles and analyse evolving spatial structures with help of a coarse
grained density $\rho$. We divide the total length of the ring into a
hundred bins and count the number of particles in each bin. The
coarse grained density is defined as the actual number of particles in
each bin divided by the average.

The time evolution of $\rho$ is shown in fig. \ref{fig:dichte1} on a linear
time scale and in fig. \ref{fig:dichte2} on a logarithmic time scale.
Several phases in the cooling process can be identified. First, the particles
start to form clusters and voids as they lose kinetic energy in collisions
(initially, when $T_B$ is small compared to the translational
energy, the coefficient of restitution is always close to $\gamma$). After
these clusters have formed, one observes collisions of clusters, forming
larger clusters.  Simultaneously the dissolution of clusters starts to set in,
the remains being sent outwards to join neighbouring clusters. The biggest
clusters and voids are seen to survive for times of order $10^4$. This complex
interaction of forming and dissolving clusters continues with a clear tendency
to form fewer and larger clusters.  Finally these large clusters dissolve to
establish the equilibrium state, i.~e. {\it equipartition} among all degrees
of freedom.

For 10000 particles it takes a time of order $10^5$ until the cooling dynamics
is finished and the equilibrium state for the larger system is reached,
whereas for 50 particles it takes only a time of order $10^3$ (see fig.
\ref{fig:cool1}).
The equilibrium state is reached only after the formation and dissolution of
essentially one final large cluster. In a smaller system the end of the
cascade of clusters of increasing size is reached earlier, simply because
there are fewer particles.

\vspace{1ex}
\noindent\textbf{Phase space}

\begin{figure}[hbt]
\parbox{9cm}{\epsfig{file=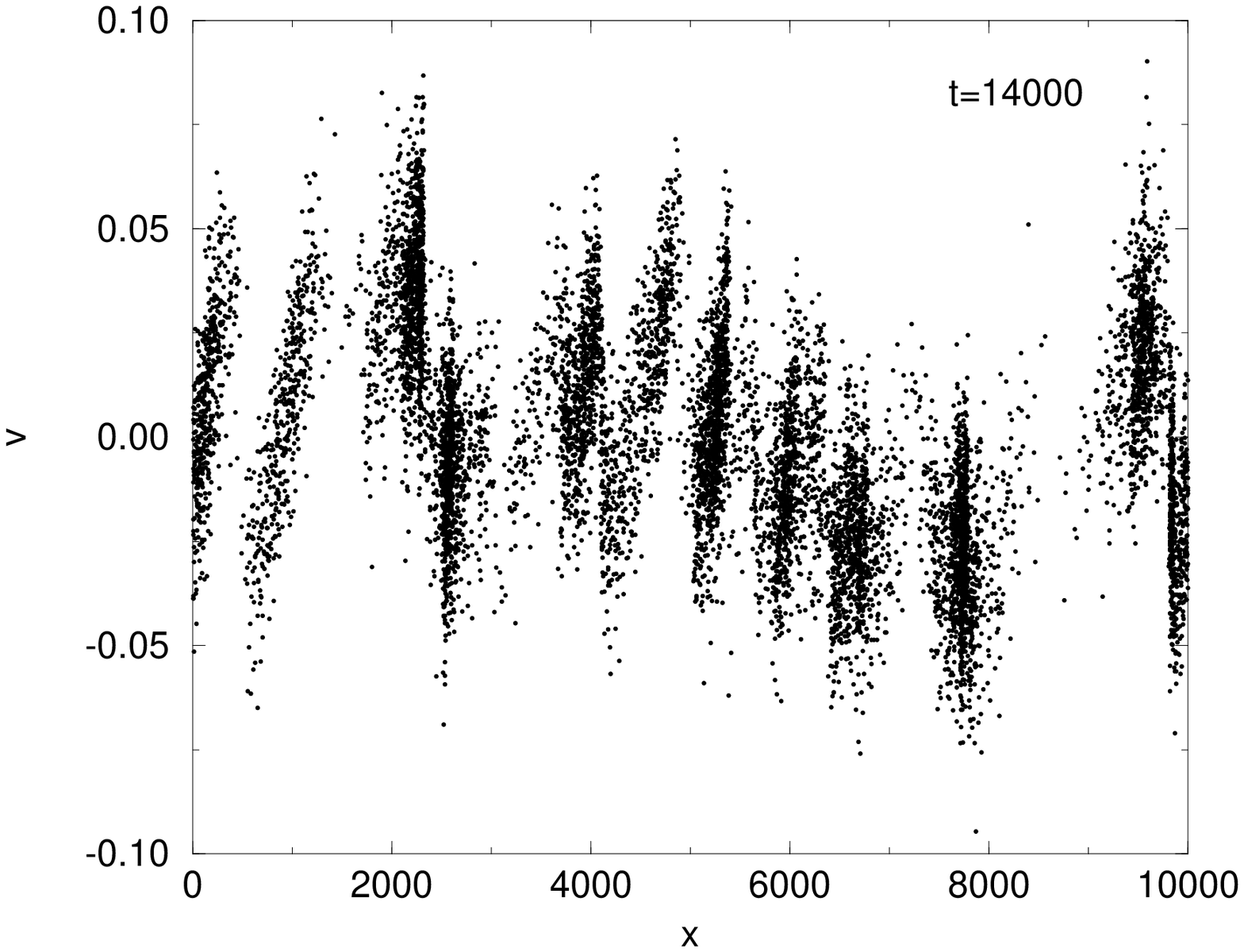,width=8.5cm}
          \raisebox{1.5cm}{\makebox[0cm][r]{(a)\hspace*{6.5cm}}}}\\
\parbox{9cm}{\epsfig{file=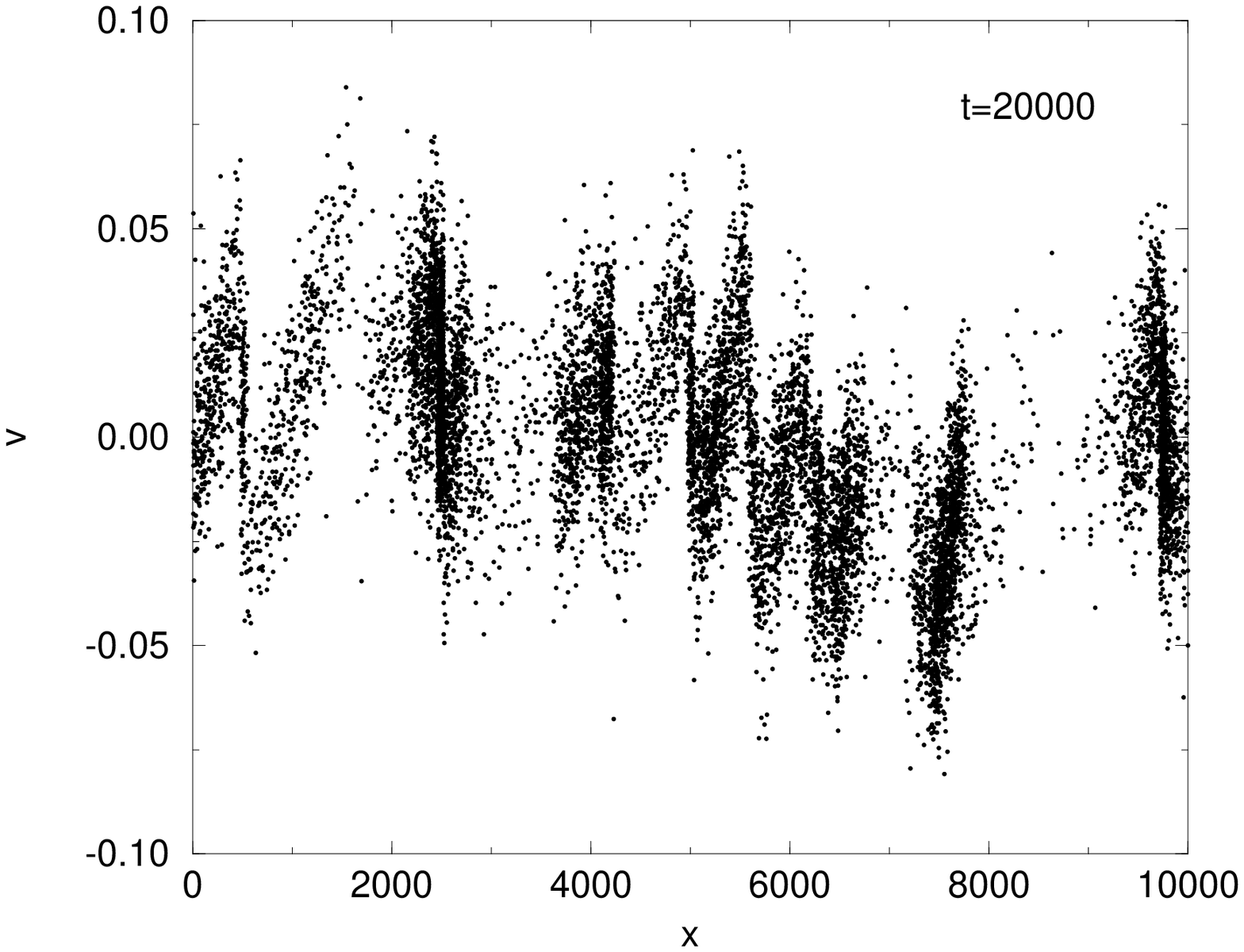,width=8.5cm}
          \raisebox{1.5cm}{\makebox[0cm][r]{(b)\hspace*{6.5cm}}}}\\
\caption{Phase space plot of the system at two different times.}
\label{fig:phase}
\end{figure}

The complete information about the state of the system at time $t$ is
contained in a phase space plot, as shown in fig. \ref{fig:phase}.  Within a
cluster of particles we expect frequent collisions and hence an effective
transfer of kinetic energy to internal vibrations.  Frequently regions of high
average density are characterized by particle velocities centered around zero.
However, we also observe clusters with an average nonzero velocity, resulting
at a later time in collisions of clusters. One such collision of two clusters
can be traced in fig. \ref{fig:phase} around $x\sim2500$. In fig
\ref{fig:phase}a (a snapshot taken at $t=14000$) one observes two clusters
both with nonzero average velocity moving towards each other, wheras in fig
\ref{fig:phase}b (taken at $t=20000$) the clusters have collided and formed a
larger one.

We also see around $x\sim 1000$ the occurrence of a stripe-shaped fluctuation
in the phase-space plot. This type of fluctuation has already been observed
and discussed by McNamara and Young \cite{Young93} and Sela and Goldhirsch
\cite{Sela95}. It gives rise to the formation of clusters out of an initially
homogeneous region. Thus figure \ref{fig:phase} shows that the dynamics of our
system are indeed rather complex as formation, movement, interaction
and dissolution of clusters all happen simultaneously.




\vspace{1ex}
\noindent\textbf{Local kinetic energy}

It is interesting to see how the kinetic energy is spatially distributed.  We
define a coarse grained kinetic energy density similar to the coarse grained
density by summing the kinetic energies of all particles inside a bin and
dividing by the number of particles in the bin.

\begin{figure}[htb]
\epsfig{file=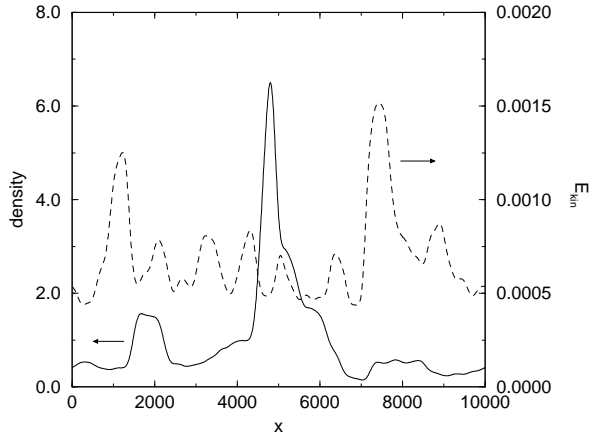,width=8.5cm}
\caption{Comparison of the local kinetic energy (dashed line) and particle
density (solid line).}
\label{fig:energy}
\end{figure}

One might be tempted to conjecture that the local kinetic energy is in some
way correlated to the clustering because most collisions occur within the
clusters. Fig. \ref{fig:energy} (as an example) reveals, however, that this is
generally not the case: although the kinetic energy shows some structure there
is no visible correlation to the density, not even in a state as the one shown
in fig. \ref{fig:energy}, where all the particles are extremely clustered.
Fig.  \ref{fig:energy} is a snapshot of the system at time $t=100343$
(\emph{cf.} figs.  \ref{fig:dichte1} and \ref{fig:dichte2}).

\vspace{1ex}
\noindent\textbf{Velocity distribution}

In the cooling stage, the system is still far from equilibrium, so
that the velocity distribution of the particles is not expected to be 
a Maxwell distribution. It is therefore interesting to test what
kind of distribution the velocities really follow.

\begin{figure}[bht]
\epsfig{file=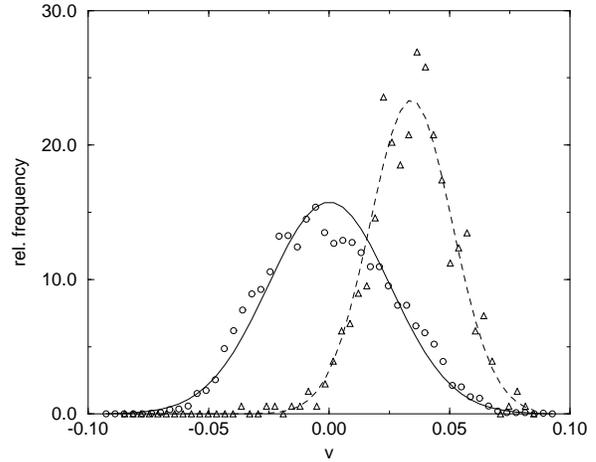,width=8.5cm}
\caption{Velocity distribution of all particles (circles) and the particles
inside one particular cluster (triangles) at time $t=14000$. The cluster
chosen for this curve is centered around $x=2000$ (\emph{cf.} figs.
\ref{fig:dichte1}, \ref{fig:dichte2} and \ref{fig:phase}a)}
\label{fig:maxwell}
\end{figure}

Data analysis shows that the velocity distribution of ${\it all}$ particles is
indeed not a Gaussian distribution (see fig. \ref{fig:maxwell}). There are
relatively large deviations especially near the maximum of the curve.  If one
restricts the data analysis to only those particles inside a single cluster,
however, one finds that the velocity distribution of these is to a much better
degree gaussian, considering that there are only about 1/10th of the total
number of particles in the cluster.  This can be well understood because there
are many collisions between particles inside a cluster and thus a local
equilibrium is reached, resulting in a Maxwellian distribution. On the other
hand the velocity distribution of all particles reflects the velocity
distribution of the \emph{clusters}. As long as the complicated process of
forming and dissolution of clusters is underway, the clusters are naturally
far away from equilibrium. This leads to the observed deviations from the
gaussian curve.

Since our system is far away from the quasielastic limit, we see quite a
different velocity distribution than MacNamara and Young \cite{Young93}, who
simulated a one dimensional system of quasielastic particles. They observed a
bimodal velocity distribution because the particles tend to concentrate on the
upper and lower edges of a band in a phase space plot similar to fig.
\ref{fig:phase}. In our simulation, the situation is much more complex
because of the formation of many clusters, each with its own velocity
distribution.

\vspace{1ex}
\noindent\textbf{Correlation function}

\begin{figure}[hbt]
\epsfig{file=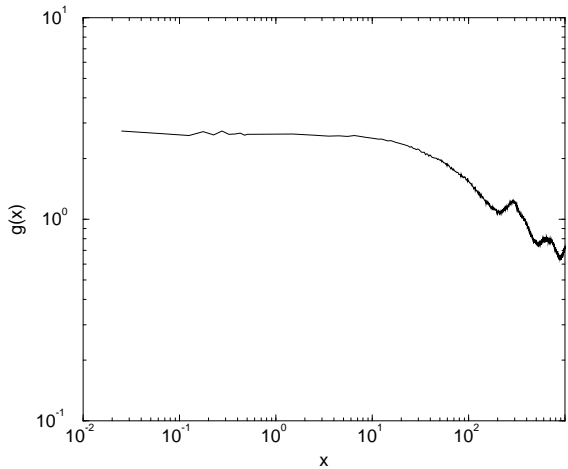,width=8.5cm}
\caption{The pair correlation function $g(x)$ of the system at $t=14000$.}
\label{fig:correlation}
\end{figure}

The inelasticity of collisions leads to a clustering of particles, as
can be seen in figs. \ref{fig:dichte1} and \ref{fig:dichte2}. Williams
\cite{Williams} has described a one dimensional system of individually heated
granular particles. He found that the pair correlation function, 
defined by $g(x)=\frac{1}{N-1}\sum_{i\ne j}\delta(x-|x_i-x_j|)$ of the system
in the steady state approximately follows a power law.  Here, we observe quite
a different behaviour of the correlation function (see fig.
\ref{fig:correlation}).
Instead of showing a divergence at zero separation, it levels off to a
plateau. The explanation for such a different behaviour lies in the mechanism
of heating: When the particles are heated individually, i.~e. when they are
driven by a random force, they will half of the time be kicked back in the
direction of the particle that they last collided with. Thus there is some
additional tendency for the particles to stick together. In our model, however,
the particles will only change their velocity when they collide, thus
favouring larger distances.

It should be noted that the correlation function in fig. \ref{fig:correlation}
is not that of the steady state of our system but a snapshot taken during
the cooling process. The steady state of our model is
trivial, implying a constant correlation function, $g(x)\equiv 1$.


\section{Damped internal modes}
\label{damping}

So far we have considered a conservative system, i.~e. the total energy of
translational motion and internal vibrations is conserved. Such a model gives
rise to a stationary equilibrium state in which equipartition among all
degrees of freedom holds, so that the translational momenta are of order
$O(1/\sqrt{N_{mod}})$. To model granular media, one should take into 
account additional
dissipative mechanisms, which result in a decrease of the total energy so that
the particles are truly at rest in the stationary state. One such mechanism is
black body radiation.

A simple way to model this effect is to let the bath temperature of
each particle decay in time. Hence we suggest the following
modification of the algorithm of sec. \ref{algorithm}. 
Inbetween collisions the
particles move freely and their bath temperature decreases according
to a simple exponential decay
\begin{equation}
T_B^i(t)=T_B^i(t_i)\exp{(-(t-t_i)\nu)} \quad\mbox{for} \quad t>t_i
\end{equation}
Here $t_i$ denotes the instant of time when the last collision of particle $i$
took place. The same decay frequency $\nu$ is used for all particles. The
updating of relative velocities and bath temperatures in a collision is
unchanged as compared to sec. \ref{algorithm}.

We expect that the effect of such a dissipative mechanism will strongly depend
on the frequency of collisions as compared to $\nu$. If collisions are
rather infrequent, then the decay will be effective and the bath of the
particles will cool down in between collisions. The resulting dynamics should
resemble the deterministic case and hence one should observe a strong increase
in the collision frequency, because the system is developing towards
inelastic collapse. When this happens, the collision frequency becomes
comparable or even smaller than the decay rate $\nu$. In that case the
internal modes can no longer relax in between collisions.  In the limit of
very high collision frequencies the bath temperatures are effectively
non-decaying, so that one recovers the algorithm of sec. \ref{algorithm}
without any dissipation.  Hence we expect to see the system develop towards
inelastic collapse with a strong increase in collision frequency, followed by
a period of time where the collision frequency levels off.

\begin{figure}[htb]
\epsfig{file=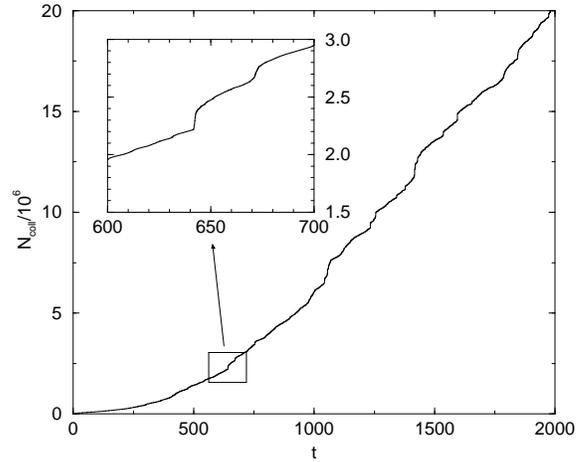,width=8.5cm}
\caption{Number of collisions as a function of time for the dissipative
  system.}
\label{fig:dissncoll}
\end{figure}

\begin{figure}[htb]
\epsfig{file=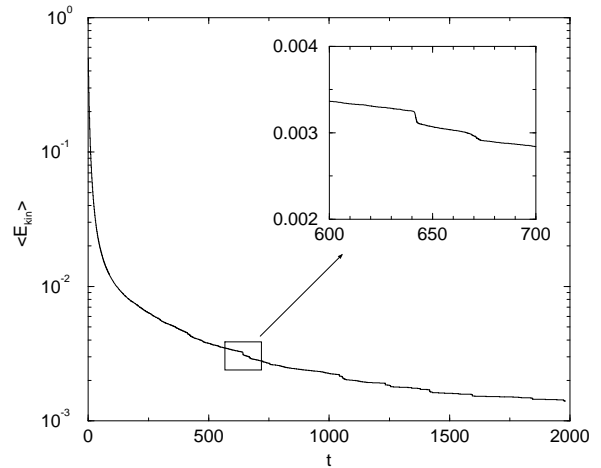,width=8.5cm}
\caption{Decrease of the total kinetic energy as a function of time for the
  dissipative system.}
\label{fig:dissenergy}
\end{figure}

These expectations are confirmed by numerical simulations of once again 10000
particles with $\nu=0.01$. In fig. \ref{fig:dissncoll} we show the total
number of collisions as a function of time.
One clearly observes rather sharp steps followed by smoother regions, as
explained above.  In fig. \ref{fig:dissenergy} we show the decrease in total
kinetic energy as a function of time. Steep regions of the collision frequency
correspond to steep regions in the energy plot because the frequent collisions
among clustered particles draw much energy out of the system.

\begin{figure}[htb]
\epsfig{file=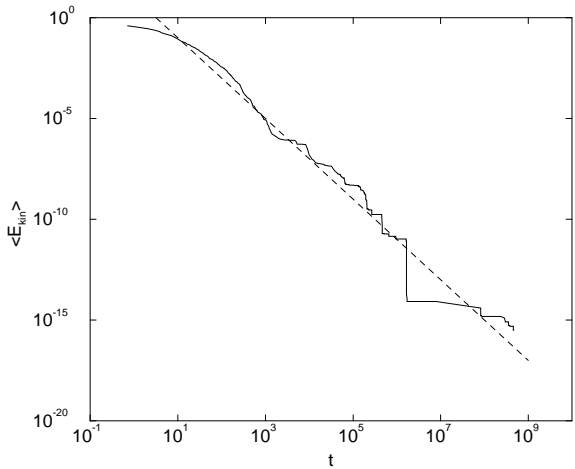,width=8.5cm}
\caption{Decrease of the kinetic energy as a function of time for 50
dissipative particles. The dashed line shows the $E\sim t^{-2}$ relation.}
\label{fig:dissfinal}
\end{figure}

An important question is the following: What is the stationary state
of the system with dissipation and how long does the system need to
relax to the stationary state? The final state should be one big
cluster, with all particles at rest.
As explained above, the dynamics with dissipation resemble the dynamics of a
deterministic system as long as no inelastic collapse is at hand. For this
reason it can be expected that the kinetic energy on the average follows the
mean field result $E\sim t^{-2}$, occasionally disrupted by the occurence of a
cluster, which is however quickly dissolved. Simulations show that this
behaviour can indeed be observed, see fig. \ref{fig:dissfinal}. Since the
above mentioned scaling law never permits the energy to become exactly 0 in a
finite time, the stationary state (which has energy 0) can also never be
reached in finite time.

\section{Conclusion}
\label{conclusion}

We have presented the results of simulations performed on a recently developed
model for a one dimensional granular medium. The model allows for an algorithm
which is a hybrid of Monte Carlo and Event Driven and which avoids the
inelastic collapse. Thus, we have been able to perform \emph{long} simulations
on a \emph{large} system far away from the quasielastic limit. The model in
its simplest form conserves energy: translational energy can be transferred to
internal vibrational modes of the particles and vice versa.  Starting from a
state with no internal modes excited, there is a long cooling regime,
extending over several orders of magnitude in time, before the stationary
state, characterized by equipartition of energy, is reached. The decrease
of kinetic energy during this cooling stage shows considerable deviations from
the mean field result $E_{kin}\sim t^{-2}$. We have also observed a complex
process of cluster forming, movement, interaction and dissolution. Inside the
clusters we find that the particles are close to local equilibrium, which is
indicated by the fact that a Maxwellian velocity distribution with in general
nonzero mean velocity holds for particles in the cluster.

The model has been extended to include net dissipation of energy by
exponential damping of the internal modes. In this case, the algorithm still
shows no inelastic collapse.  On the average, the decrease of the kinetic
energy follows the mean field result but with considerable fluctuations due to
the complex cluster dynamics which are still a feature of the model with
energy dissipation.

Thus our model, which is based on a \emph{microscopic} mechanism for the loss
of translational energy during collisions, is well suited as a starting point
for simulation and theoretical description of one dimensional granular media.
Unlike many other models, it makes use of an exact treatment of the collision
dynamics of the colliding rods and hence offers a possible intuitive way of
understanding the precise manner in which translational energy is removed from
a granular system. Future work will use this model to investigate the 
properties of driven granular assemblies. 
In our model a specific mechanism -- transfer of translational energy to
internal vibrations -- has been analysed to develop a microscopic basis for an
effective coefficient of restitution. One may wonder which of our results
depend on the particular mechanism. To study this question, we are presently
investigating distributions $p_{\beta}(\epsilon)$ which are only restricted
by detailed balance and not derived from a microscopic model. One may
also try to extend our analysis to higher dimensional objects like disks
or spheres. In the simplest geometry these objects are colliding in a one
dimensional tube, so that no tangential forces like coulomb friction have
to be considered. Our microscopic model is easily generalized to this case
and allows for investigating how effectively energy of translation is 
transferred to elastic vibrations \cite{franz}. The frequently used
quasistatic approximation of Hertz implies no energy transfer at all.



\begin{thebibliography}{99}
\bibitem{GieseAZ96}
G. Giese and A. Zippelius, Phys. Rev. E{\bf 54}, 4828 (1996).
A short summary of the results has been published in \cite{Giese96}.
\bibitem{Giese96}
G. Giese  in D. E. Wolf, M. Schreckenberg, and A. Bachem (eds.):
{\em Traffic and Granular Flow} (World Scientific, Singapore 1996), p.~335.
\bibitem{Bernu90}
B. Bernu and R. Mazighi, J. Phys. A {\bf 23}, 5745 (1990);
\bibitem{Young93}
S. McNamara and W. R. Young, Phys. Fluids {\bf 4}, 496 (1992) and
S. McNamara and W. R. Young, Phys. Fluids {\bf 5}, 34 (1993).
\bibitem{Sela95}
N. Sela and I. Goldhirsch, Phys. Fluids {\bf 7}, 507 (1995).
\bibitem{Clement93}
E. Clement, S. Luding, A. Blumen, J. Rajchenbach and J. Duran,
Intern. J. Mod. Phys. B{\bf 7}, 1807 (1993);
\bibitem{Luding94}
S. Luding, E. Clement, A. Blumen, J. Rajchenbach, and J. Duran,
Phys. Rev. E{\bf 49}, 1634 (1994).
\bibitem{Luck93}
J. M. Luck and Anita Mehta, Phys. Rev. E{\bf 48}, 3988 (1993).
\bibitem{Bridges88}
A. P. Hatzes, F. G. Bridges, and D. N. C. Lin,
Mon. Not. R. astr. Soc. \textbf{231}, 1091 (1988);
K. D. Supulver, F. G. Bridges, and D. N. C. Lin,
Icarus \textbf{113}, 188 (1995)
\bibitem{Hertz1882}
H. Hertz, J. Reine Angew. Math. {\bf 92}, 156 (1996).
\bibitem{Poeschl28}
Th. Poeschl, Z. Phys. {\bf 46}, 142 (1928).
\bibitem{Pao55}
Y.- H. Pao, J. Appl. Phys. {\bf 26}, 1083 (1955).
\bibitem{Brilliantov96}
N. V. Brilliantov, F. Spahn, J. M. Hertzsch, and T. Poeschel,
Phys. Rev. E{\bf 53}, 5382 (1996); 
J. M. Hertzsch, F. Spahn, and N. V. Brilliantov,
J. Phys. (Paris) II, {\bf 5}, 1725 (1995).
\bibitem{Spahn97}
J. Hertzsch, H. Scholl, F. Spahn, and I. Katzorke, Astron. \& Astrophys. 
{\bf 320}, 319 (1997);
F. Spahn, U. Schwarz, and J. Kurths, Phys. Rev. Lett. {\bf 78}, 1596 (1997).
\bibitem{Auerbach}
For a pedagogical discussion of the wave theory of elastic rods see e.g.
D. Auerbach, Am. J. Phys. {\bf 62}, 522 (1994).
\bibitem{Kloeden}
see e.g. P. E. Kloeden and E. Platen, {\em Numerical solution of
  stochastic differential equations} (Springer, Berlin, 1992), p.~42.
\bibitem{tobepublished}
T. Aspelmeier and A. Zippelius, \emph{to be published}.
\bibitem{Williams}
D. R. M. Williams, Physica A{\bf 233}, 718 (1996).
\bibitem{one.d.sim}
Y. Du, H. Li, and L. P. Kadanoff, Phys. Rev. Lett. {\bf 74}, 1268 (1995);
R. Mazighi, B. Bernu, and F. Delyon,Phys. Rev. E{\bf 50}, 4551 (1994);
\bibitem{franz}
F. Gerl, \emph{private communication}

\end{thebibliography}
\end{document}